\documentclass[aps,pra,twocolumn,showpacs,epsfig]{revtex4}
\usepackage{dcolumn}
\usepackage{bm}
\usepackage{graphicx}
\usepackage{amsmath}
\usepackage{latexsym}
\usepackage{amsfonts}
\usepackage{amssymb}
\usepackage{array}
\usepackage{times}
\usepackage{epsfig}
\usepackage{epstopdf}
\usepackage{setspace}

\newcommand{\ket}[1]{\left\vert#1\right\rangle}

\newcommand{\bra}[1]{\left\langle#1\right\vert}

\begin{document}
\title{Memory-keeping effects and forgetfulness in the dynamics of a qubit coupled to a spin chain}
\author{Tony J. G. Apollaro$^1$, Carlo Di Franco$^2$, Francesco Plastina$^3$, and Mauro Paternostro$^4$}
\affiliation{$^1$Dipartimento di Fisica e Astronomia, Universit\`a di Firenze, Via G. Sansone 1, I-50019 Sesto Fiorentino (FI), Italy\\
$^2$Department of Physics, University College Cork, Republic of Ireland\\
$^3$Dipartimento di  Fisica, Universit\`a della Calabria,  87036 Arcavacata di Rende (CS), Italy \&
 INFN - Gruppo collegato di Cosenza \\
$^4$School of Mathematics and Physics, Queen's University, Belfast BT7 1NN, United Kingdom }

\newcommand{\nbar}{\overline{n}}
\newcommand {\mean}[1]{\langle #1 \rangle}
\newcommand {\meantwo}[1]{\langle #1 \rangle}
\newcommand{\parall}{\uparrow\uparrow}
\newcommand{\antiparall}{\uparrow\downarrow}
\newcommand\lavg{\left\langle}
\newcommand\ravg{\right\rangle}

\begin{abstract}
Using recently proposed measures for non-Markovianity [H. P.
Breuer, E. M. Laine, and J. Piilo, Phys. Rev. Lett. {\bf 103},
210401 (2009)], we study the dynamics of a qubit coupled to a 
spin environment via an energy-exchange mechanism. We
show the existence of a point, in the parameter space of the
system, where the qubit dynamics is effectively Markovian and that
such a point separates two regions with completely different
dynamical behaviors. Indeed, our study demonstrates that the qubit
evolution can in principle be tuned from a perfectly forgetful one
to a deep non-Markovian regime where the qubit is strongly
affected by the dynamical back-action of the environmental spins.
By means of a theoretical quantum process tomography analysis, we
provide a complete and intuitive characterization of the qubit
channel.
\end{abstract}

 \maketitle

In the study of open quantum systems, quite often the Markovian
approximation has been a useful starting point to describe their
dynamics. 
Only very recently, new and powerful tools have been
designed in order to tackle the important question of explicitly
quantifying the non-Markovian character of a 
system-environment interaction or a dynamical map
\cite{breuer09,rivas09,wolf08}. Such a task is extremely
important, given that non-Markovian effects are known or expected
to occur in a wide range of physical situations, especially in the
realm of solid-state devices where a system of interest is often
exposed to memory-preserving environmental mechanisms. As a specific example,
switching impurities have been shown to affect superconducting
devices in various regimes~\cite{martinis}.

From a fundamental point of view, devising a reliable way to
actually quantify non-Markovianity is very useful in light of the
plethora of frequently {\it ad hoc} or technically rather involved
approaches put forward so far in order to study memory-keeping
environmental actions. A few instances emerge as promising
measures for the non-Markovian nature. Wolf {\it et
al.}~\cite{wolf08} have proposed to quantify the degree of
non-Markovianity of a map by considering the minimum amount of
noise required in order to make the evolution of a system fully
Markovian. On the other hand, in Ref.~\cite{rivas09}, Rivas and
co-investigators discussed two approaches, founded on the
deviations of a given map from full divisibility. Finally, Breuer
{\it et al.} proposed a way to quantify non-Markovian effects by
looking at the back-action induced on the system under scrutiny by
its memory-keeping environment~\cite{breuer09}. Further approaches
have been considered and measures have been proposed which are
based on the use of other interesting instruments such as the
quantum Fisher information~\cite{sun09}.

In this paper, we focus on the non-Markovianity measure proposed
in Ref.~\cite{breuer09} to study the dynamics of a qubit coupled
to a spin environment described by an $XY$ model in a transverse
magnetic field. Our aim is to analyze a simple and yet non-trivial
system-environment set, displaying a broad range of behaviors in
the parameter space, in order to relate the features of the
non-Markovianity measure to the known spectral and dynamical
structures of the spin system. The choice of the model has been
also dictated by the versatility demonstrated by spin-network
systems in the engineering of protocols for short-range
communication~\cite{boseetal} and also for the investigation of
the interplay between quantum-statistical and
quantum-information-related aspects~\cite{rmp}.

Intuitively, one would expect that the coupling to a spin environment
always leads to a non-Markovian dynamics for the qubit. This can be justified 
by noticing that the environmental correlation time is non-zero even in the weak 
coupling regime and by conjecturing that, by increasing the coupling strength, a
Markovian approximation can only become less valid. However, we
demonstrate that this is not at all the case and that peculiar
behaviors occur at intermediate couplings. 
The chosen measure for non-Markovianity, in fact, turns out to be
identically null at specific dynamical regimes, thus
demonstrating the absence of a net {\it re-flux} of information
from the environment back to the qubit. Our study of
non-Markovianity is thus novel and makes the use
of measures for its quantification  a valuable tool for the
identification of {\it special conditions} in the parameter space
of the environment. Moreover, our study opens up the possibility
to exploit the different memory-keeping regimes induced by
controlling and tuning the properties of the environment to
effectively ``drive" the qubit dynamics in a non-trivial and
potentially very interesting way. The effective evolution of our
two-level system could be guided across different regimes, ranging
from strong environmental back-action to completely forgetful
dynamics typical of a Markovian map. The potential of such
flexibility for reliable control at the quantum level is a topic
that will be explored in future.

The reminder of this paper is organized as follows. In
Sec.~\ref{modelmeasure} we introduce the model under scrutiny and
revise the basic principles behind the chosen measure for
non-Markovianity. We study a few cases amenable to a full analytical
solution and highlight how, in a few of such instances, a simple
experimental protocol can be designed for the inference of the
properties of the qubit dynamics. Most importantly, we reveal the
existence of an operating regime where the chosen measure is
strictly null and relate such an effect to intriguing
modifications occurring at the level of the energy spectrum of the
qubit-environment system. Sec.~\ref{caratterizzazione} is devoted
to the formal characterization of such peculiar point in the
parameter space. We first demonstrate divisibility of the
corresponding dynamical map, hence its Markovian nature, and then
perform a theoretical analysis based on the use of
 quantum process tomography to quantitatively infer its
properties. Finally, in Sec.~\ref{conc} we draw our conclusions.

\section{The model and the measure}
\label{modelmeasure}

We consider a qubit $Q$ coupled to a chain $\Gamma$ of $N$
interacting spin-$1/2$ particles. The qubit is described by the
spin-$1/2$ vector operator $\hat{\bm s}_{0}$, while the operator
$\hat{\bm s}_{n}$ (${n{=}1,..,N}$) corresponds to the spin located
at site $n$ of the chain $\Gamma$. The logical basis for the spins
and $Q$ is given by $\{\ket{0}_j,\ket{1}_j\}$ with $j=0,..,N$. The
Hamiltonian ruling the intra-chain interaction is taken to be of
the $XY$-Heisenberg type (we set $\hbar=1$)
\begin{equation}
\hat{\cal{H}}_{\Gamma}{=}-2\!\sum_{n=1}^{N-1}
(J^x_{n} \hat{s}^x_{n} \hat{s}^x_{n+1}\!
+\!J^y_{n} \hat{s}^y_{n} \hat{s}^y_{n+1})
-2\sum_{n=1}^{N} h^{}_{n}\hat{s}_{n}^z,
\label{e.H_Gamma}
\end{equation}
where $h_{n}$ is the local field applied at site $n$. $\Gamma$ is
open-ended with $J^{x,y}_{n}$'s and $h_{n}$'s being not
necessarily uniform. The qubit is coupled to the first spin of the
environment, embodied by $\Gamma$, via an exchange interaction  of
strengths $J^{x,y}_{0}$ and is subjected to a local field
${h}_{0}$ according to
\begin{equation}
\hat{\cal{H}}_{0}\!=\!-2(
J^x_{0}\hat{s}^x_{0}\hat{s}^x_{1}+
J^y_{0}\hat{s}^y_{0}\hat{s}^y_{1})\!-\!
2h_{0}\hat{s}^z_{0}.
\label{e.H_qubit}
\end{equation}
In order to determine the time-evolution of $Q$, we resort to the
Heisenberg picture and the formal apparatus put forward in
Refs.~\cite{DiFrancoEtal07}, which provides particularly powerful
tools for the study of the many-body problem embodied by
$\hat{\cal H}_0+\hat{\cal H}_\Gamma$. Using the operator-expansion
theorem and the algebra satisfied by Pauli matrices, we find that
the time evolution of the components of ${\hat{\bm s}}_0$ reads
\begin{equation}
\label{e.XYZ0(t)}
\begin{aligned}
\hat{s}^x_0(t)&=\frac{1}{2}\sum_{n=0}^N
\left[\Pi^x_n(t)\hat{\sigma}^x_n+\Delta^x_n(t)\hat{\sigma}^y_n\right]\hat P_n~,
\\
\hat{s}^y_0(t)&=\frac{1}{2}\sum_{n=0}^N
\left[\Pi^y_n(t)\hat{\sigma}^y_n-\Delta^y_n(t)\hat{\sigma}^x_n\right]\hat P_n~,
\\
\hat{s}^z_0(t)&=-i\hat{\sigma}^x_0(t)\hat{\sigma}^y_0(t)/2,
\end{aligned}
\end{equation}
where $\hat{\sigma}_n^\alpha$ ($\alpha{=}x,y,z$) are the Pauli
operators for the spin at site $n$ and
$\hat P_n=\prod_{i=1}^{n-1}\hat{\sigma}^z_i$.
The time-dependent coefficients $\Pi_n^{x}(t)$
and $\Delta_n^{x}(t)$ are the components of the $(N+1)$-dimensional
vectors ${\bm\Pi}^{x}(t)$ and ${\bm\Delta}^{x}(t)$ defined by
\begin{eqnarray}
{\bm\Pi}^x(t)&=&\sum_{p=0}^{\infty}(-1)^p\frac{t^{2p}}{(2p)!}
({\bm \tau}{\bm \tau}^T)^p{\bm v},
\label{e.PiX}\\
{\bm\Delta}^x(t)&=&\sum_{p=0}^{\infty}(-1)^{p}\frac{t^{2p+1}}{(2p+1)!}
{\bm \tau^T}({\bm \tau}{\bm \tau}^{T})^p{\bm v},
\label{e.DeltaX}
\end{eqnarray}
where $T$ stands for transposition, the vector ${\bm v}$ has components
$v_i=\delta_{i0}$ and
the tri-diagonal adjacency matrix ${\bm \tau}$ has elements
\begin{equation}
\tau_{ij}=
J_{i-1}^x\delta_{i-1,j}+J_i^y\delta_{i+1,j}-2 h_i\delta_{i,j}.
\label{e.T}
\end{equation}
Notice that we have labelled columns and rows of
$(N{+}1){\times}(N{+}1)$ matrices and $(N{+}1)$-dimensional
vectors using indices ranging from $0$ to $N$. The coefficients
${\bm \Pi}^{y}(t)$ and ${\bm \Delta}^{y}(t)$ are obtained from
Eqs.~(\ref{e.PiX}) and (\ref{e.DeltaX}) by replacing ${\bm \tau}$
with ${\bm \tau}^T$. Both $\bm{\tau\tau}^T$ and
$\bm{\tau}^T\bm\tau$ can be easily diagonalized by orthogonal
matrices ${\bf U}$ and ${\bf V}$ such that
$({\bm\tau}{\bm\tau}^T)^p{=}{\bf U}{\bm\Lambda}^{2p}{\bf U}^T$ and
$({\bm\tau}^T{\bm\tau})^p{=}{\bf V}{\bm\Lambda}^{2p}{\bf V}^T$,
with ${\bm\Lambda}$ a diagonal matrix whose elements
$\lambda_{ij}=\lambda_i\delta_{ij}$ are the (positive) square
roots of the eigenvalues of $\bm{\tau}^T\bm\tau$. Consequently,
Eqs.~(\ref{e.PiX}) and (\ref{e.DeltaX}) can be fully summed up to
give
\begin{equation}
\label{e.PiDeltaxyResum}
\begin{aligned}
{\bm \Pi}^x(t)&={\bm U}{\bm \Omega}(t){\bm U}^T{\bm v},~~
{\bm \Delta}^x(t)={\bm V}{\bm\Sigma}(t){\bm U}^T{\bm v},\\
{\bm \Pi}^y(t)&={\bm V}{\bm \Omega}(t){\bm V}^T{\bm v},~~
{\bm \Delta}^y(t)={\bm U}{\bm\Sigma}(t){\bm V}^T{\bm v},
\end{aligned}
\end{equation}
where ${\bm \Omega}(t)$ and ${\bm \Sigma}(t)$ are diagonal
matrices with elements $\Omega_{ij}(t){=}\cos(\lambda_i t)
\delta_{ij}$ and $\Sigma_{ij}(t){=}\sin(\lambda_i t) \delta_{ij}$.

By using Eqs.~(\ref{e.XYZ0(t)}), one can determine the time evolution
of the state of $Q$ as
${\rho}(t){=}\hat{\openone}/2{+}\sum_{\alpha}
\mean{\hat{s}^\alpha_0(t)}\hat{\sigma}^\alpha_0$. In the
evaluation of the expectation values required to determine
$\rho(t)$ we assume that $Q$ and $\Gamma$ are initially
uncorrelated. The conservation rule
${[\hat{\cal{H}},\bigotimes_{n=0}^{N}\hat\sigma^z_{n}]=0}$ and the
property $\mean{\hat\sigma_n^\alpha\hat\sigma_{m\neq
n}^{\beta\neq\alpha}}{=}0$ imply
\begin{equation}
\label{dinamica}
\begin{aligned}
&\mean{\hat s^x_0(t)}{=}\frac{1}{2}\sum_{n=0}^N
\left[\Pi^x_n(t)\mean{\hat P_n\hat\sigma^x_n}
+\Delta^x_n(t)\mean{\hat P_n\sigma^y_n}\right],\\
&\mean{\hat s^y_0(t)}{=}\frac{1}{2}\sum_{n=0}^N
\left[\Pi^y_n(t)\mean{\hat P_n\hat\sigma^y_n}-
\Delta^y_n(t)\mean{\hat P_n\hat\sigma^x_n}\right],\\
&\mean{\hat s^z_0(t)}{=} \frac{1}{2}\sum_{n=0}^N
\left[\Pi_n^x(t)\Pi_n^y(t) + \Delta_n^x(t)\Delta_n^y(t)\right]
\mean{\hat\sigma_n^z}\\
-&\frac{1}{2}\sum_{n<m}^N
\left[\Pi_n^y(t) \Pi_m^x(t) + \Delta_n^x(t) \Delta_m^y(t)\right]
\mean{\hat P_{n+1}\hat P_m\hat\sigma_n^x\hat\sigma_m^x}\\
-&\frac{1}{2}\sum_{n<m}^N
\left[\Pi_n^x(t) \Pi_m^y(t) + \Delta_n^y(t) \Delta_m^x(t)\right]
\mean{\hat P_{n+1}\hat P_m\hat\sigma_n^y\hat\sigma_m^y}.
\end{aligned}
\end{equation}
In order to evaluate the above equations one needs multi-spin
correlation functions, involving, in particular, the degrees of
freedom of $\Gamma$. To this end, we consider it to be in its
ground state in what follows. However, as it will be shown later
on, the value of the measure of non-Markovianity chosen for this
work is independent of the state of $\Gamma$, provided that
$\mean{s^{x(y)}_n(0)}=0$ holds, as it does for the ground state with no broken symmetry.

By using Eqs.~(\ref{dinamica}), one can finally determine ${\rho}(t)$.
The interaction with the spin chain acts for $Q$ as a dynamical
map $\Phi(t,0)$ such that $\rho(t)=\Phi(t,0)\rho(0)$. The
properties of the map depend on the relative weight of the various
parameters entering $\hat{\cal H}_0{+}\hat{\cal H}_\Gamma$. Our
aim is to characterize the non-Markovian nature of $\Phi$ as a
function of these parameters.

To pursue our task, we consider the measure proposed in
Ref.~\cite{breuer09}, based on the study of the time-behavior of
the trace distance ${\cal
D}[{\rho}^{(1)}(t),{\rho}^{(2)}(t)]{=}\frac{1}{2}\text{Tr}|{\rho}^{(1)}(t){-}{\rho}^{(2)}(t)|{\in}[0,1]$
between two single-qubit density matrices ${\rho}^{(1,2)}(t)$. The
trace distance is such that ${\cal
D}[{\rho}^{(1)}(t),{\rho}^{(2)}(t)]{=}1$ when the two probed
states are completely distinguishable, while it gives $0$ for
identical states~\cite{NC}. The degree of non-Markovianity ${\cal
N}(\Phi)$ of the dynamical map $\Phi$, is defined as
\begin{equation}
{\cal N}(\Phi){=}\max\!\sum_n\{{\cal D}[{\rho^{(1)}(b_n)},{\rho^{(2)}(b_n)}]{-}{\cal D}[\rho^{(1)}(a_n),\rho^{(2)}(a_n)]\}
\end{equation}
where the maximization is performed over the states ${\rho^{(1,2)}(0)}$
and $(a_n,b_n)$ is the $n^{\text{th}}$ time window such that
\begin{equation}
\sigma[t,\rho^{(1,2)}(0)]=\partial_t{\cal D}[\rho^{(1)}(t),\rho^{(2)}(t)]>0.
\end{equation}
The function $\sigma[t,\rho^{(1,2)}(0)]$, which has been dubbed
{\it flux of information} in Ref.~\cite{breuer09}, encompasses {\it per
se} the condition for revealing non-Markovianity of an evolution:
the mere existence of even a single region where
$\sigma[t,\rho^{(1,2)}(0)]{>}0$ is sufficient to guarantee the
non-Markovian nature. Conceptually, in fact, ${\cal N}(\Phi)$
accounts for all the temporal regions where the distance between
two arbitrary input states increases, thus witnessing a re-flux of
information from the environment to the system under scrutiny.
Such re-flux of information amplifies the difference between two
arbitrarily picked input states evolved up to the same instant of
time. A Markovian dynamics is such that the above-mentioned
re-flux never occurs and $\sigma[t,\rho^{(1,2)}(0)]<0$ always.
For the case at hand and for two generic input density matrices
$\rho^{(1,2)}(0)$, we find
$D[{\rho}^{(1)}(t),{\rho}^{(2)}(t)]=\sqrt{\zeta}$,
where
\begin{equation}
\label{tracexy}
\begin{aligned}
\zeta&=\left[\Pi_0^x(t)\Pi_0^y(t)+\Delta_0^x(t)\Delta_0^y(t)\right]^2 p^2\\
&+\left|c f_+(\Pi_0,\Delta_0,t)+c^*f_-(\Pi_0,\Delta_0,t)\right|^2/{4}
\end{aligned}
\end{equation}
with $f_\pm(\Pi_0,\Delta_0,t){=}\Pi_0^x(t)\pm\Pi_0^y(t){\pm }i
\left[\Delta_0^x(t){\pm}\Delta_0^y(t)\right]$ where
$p{=}\rho^{(1)}_{11}(0){-} \rho^{(2)}_{11}(0)$ and
$c{=}\rho^{(1)}_{01}(0){-}\rho^{(2)}_{01}(0)$. We have used the
notation $\rho^{(a)}_{ij}(0){=}{}_0\langle{i}|\rho^{(a)}\ket{j}_0$
with ${i,j}{=}0,1$ and $a{=}1,2$. It is worth noticing that in the
above equation the initial state of the environment is completely
absent, so that the environment's multi-spin correlators are not
relevant. Eq.~(\ref{tracexy}) can be recast into a much more intuitive form
by referring to the Bloch vectors
$\textbf{r}_a{=}{(r^x_a,r^y_a,r^z_a)}$ representative of the state
$\rho^{(a)}$ (we have chosen $r^\alpha=\mean{\hat s_0^\alpha}$ as
the mean value of the spin-${1}/{2}$ operator rather than that
of the more usual Pauli operator to avoid the appearance of
irrelevant ${1}/{2}$ factors). By calling $\Delta
r^\alpha{=}r^\alpha_1{-}r^\alpha_2$ the difference between the
vectors of two input states, we have
\begin{equation}
\label{tracexy1}
\begin{aligned}
\zeta&{=}\left|\textbf{r}_1(t){-}\textbf{r}_2\right(t)|^2{\equiv}[\Pi_0^x(t)\Pi_0^y(t){+}\Delta_0^x(t)\Delta_0^y(t)]^2\!\left(\Delta r^z\right)^2\\
&{+}[\Pi_0^x(t) \Delta r^x{+}\Delta_0^x(t) \Delta r^y]^2{+}[\Pi_0^y(t) \Delta r^y{-}\Delta_0^y(t) \Delta r^x]^2.
\end{aligned}
\end{equation}
One can thus write the flux of information between the qubit and
the environmental chain $\Gamma$ as
\begin{equation}
 \sigma[t,\rho^{(1,2)}(0)]=\frac{\left(\Delta r^z\right)^2 A \partial_tA+B\partial_tB+C\partial_tC}{{\cal D}[\rho^{(1)}(t),\rho^{(2)}(t)]},
\end{equation}
where $A=\Pi_0^x(t)\Pi_0^y(t)+\Delta_0^x(t)\Delta_0^y(t)$,
$B=\Pi_0^x(t) \Delta r^x+\Delta_0^x(t) \Delta r^y$, and
$C=\Pi_0^y(t) \Delta r^y-\Delta_0^y(t) \Delta r^x$. Due to the
non-negativity of ${{\cal D}[\rho^{(1)}(t),\rho^{(2)}(t)]}$, the
condition for non-Markovian dynamics can be simply stated as
$\left(\Delta r^z\right)^2{A
\partial_tA+B\partial_tB+C\partial_tC}{>}0$. Numerically, it turns
out that the maximum in the corresponding measure of
non-Markovianity is achieved for $\rho^{(1,2)}(0)$ being antipodal
pure states lying on the equatorial plane of $Q$'s Bloch sphere
(we will come back to this point later in this paper). This
considerably simplifies the necessary condition for memory-keeping
dynamics to the form $\partial_t(B^2{+}C^2){>}0$.

Although our analysis can be carried out without major
complications in the general case, in order to simplify the
presentation, from now on we restrict ourselves to the case of a
uniform spin environment with equal isotropic couplings between
every pairs of nearest neighboring spins ($XX$-model) and set this
coupling constant as our energy (and inverse time) unit. We thus 
consider $J_0^x{=}J_0^y{=}J_0$, a condition under which the measure
of non-Markovianity becomes
\begin{equation}
\label{tracedistancexx}
{\cal D}[{\rho}^{(1)}(t),{\rho}^{(2)}(t)]
{=}\sqrt{(p^2 f(t) +\left|c\right|^2) f(t)},
\end{equation}
where $f(t)=\Pi^2_0(t)+\Delta^2_0(t)\in[0,1]$. The corresponding rate of change of the trace distance is
\begin{equation}
\label{ratexx}
 \sigma[t,\rho^{(1,2)}(0)]=\frac{\left(2 p^2 f(t)+|c|^2\right) f'(t)}{2 {\cal D}[\rho^{(1)}(t),\rho^{(2)}(t)]}.
\end{equation}
As $f(t){\ge}{0}$, the sign of $\sigma[t,\rho^{(1,2)}(0)]$ is
determined by $f'(t)$, regardless of the pair of input density
matrices $\rho^{(1,2)}(0)$. Yet, we should look for the states
maximizing the contributions to the trace distance within these
time intervals. As $|f(t)|{\leq}1$, such optimization is achieved
for $c{=}1$ and $p{=}0$; that is, antipodal pure states on the
equatorial plane of the Bloch sphere, in line with the (more
general) numerical findings reported above. The condition for
non-Markovianity can be further elaborated as
$f'(t){=}2[\Pi_0(t)\Pi'_0(t)+\Delta_0(t)\Delta'_0(t)]{>}0$.

To pursue the task of evaluating ${\cal N}(\Phi)$ in the quite
rich parameter space of our model, we start by considering the
case of a qubit resonant with the spin environment, and assess the
special case of $h_0{=}h$ first. It can be shown analytically that
in this case one gets ${\Pi_0(t)={{\cal J}_1(2t)}\cos(2 h t)}/{t}$
and ${\Delta_0(t)=-{{\cal J}_1(2t)}\sin(2 h
t)}/{t}$~\cite{ApollaroEtal10}, where ${\cal J}_n(x)$ is the
Bessel function of order $n$ and argument $x$. Correspondingly,
$f(t){=}{{\cal J}^2_1(2t)}/{t}^2$ and $f'(t)=-{4 {\cal
J}_1(2t){\cal J}_2(2t)}/{t^2}$. For those states with $|c|=1$ and
$p=0$ which maximize ${\cal N}$, we have the flux
\begin{equation}
\label{ratexxomo} \sigma[t,\rho^{(2,3)}(0)]=
-(2/{t}) \text{sgn}[{\cal J}_1(2t)]{\cal J}_2(2t)
\end{equation}
with $\text{sgn}[x]$ being the sign function.
Eq.~(\ref{ratexxomo}) is independent of $h$, which is joint result
of the condition $p=0$ and the invariance of the trace distance
under the global unitary transformation embodied by the operator
$\exp[{-i t h\sum_{i=0}^N{\hat \sigma_i^z}}]$ needed in order to
pass to the interaction picture. The time windows where
$\sigma[t,\rho^{(1,2)}(0)]{>}0$ are determined by the chain-rule
of the zeros of the Bessel functions. Overall, we get that the
flux is positive for a time
$t_{\sigma>0}{=}{\cup}(t^{1}_i,t^{2}_i)$ where $t^{1}_i$
($t^{2}_i$) is the $i^\text{th}$ zero of the Bessel function of order 1
(2). From this special case and the general considerations
reported above, we learn that ${\cal N}(\Phi)$ only depends on the
detuning $\delta h=h-h_0$.

Due to this fact, the simple case in which all of the
magnetic fields are absent ($h=h_0=0$), which can be tested
experimentally in an easy way, allows to draw some interesting
conclusions on the more general case $\delta h=0$. In absence of
magnetic fields, indeed, the behavior of $\sigma$ and ${\cal N}$
can be extracted directly by monitoring the dynamics of the qubit.
To show that this is indeed the case we start by noticing that, if
$h/J=h_0/J=0$, we have $\Delta_0(t){=}0$~\cite{ApollaroEtal10}, so
that both the trace distance and the flux of information are
determined by $f(t) = \Pi^2_0(t)$, which represents the squared
length of the Bloch vector, $|r(t)|^2$. This implies that, in
order for $Q$ to experience a re-flux of information from
$\Gamma$, $\Pi^2_0(t)$ has to be a strictly non-monotonic
function. Therefore, we observe non-Markovian dynamics when the
Bloch vector of the qubit is alternatively shrunk and elongated
during its evolution. This can be witnessed by reconstructing the
density matrix of $Q$ using standard quantum state tomography
techniques~\cite{statetomo,NC}, which are routinely
implemented in a variety of physical setups. However, there is
also an interesting alternative which does not require full state
reconstruction. From the first of Eqs.~(\ref{dinamica}), we get
that $\mean{\hat
s_0^x(t)}{=}\frac{1}{2}\Pi_0(t)\mean{\hat\sigma_0^x(0)}$. Thus, in
the spirit of the proposals put forward in
Refs.~\cite{DiFrancoEtal07,DiFrancoEtal08}, by preparing the state
of $Q$ in one of the eigenstates of $\hat{\sigma}^x_0$ [a choice
that would be perfectly consistent with our results on the input
states to be used for the calculation of ${\cal N}(\Phi)$] we can
measure $\langle\hat{s}^x_0(t)\rangle$ to determine the
non-Markovianity of the qubit dynamics, which would be revealed by
its non-monotonic time behavior.

\begin{figure}[t]
\includegraphics[width=0.8\linewidth]{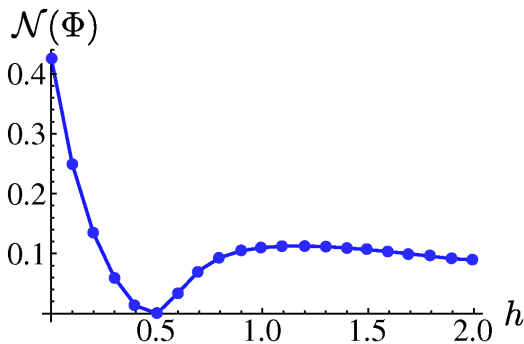}
\caption{(Color online) ${\cal N}(\Phi)$ versus $h/J$ with $h_0/J=0$,
$J_0/J=1$ and $N=100$. To avoid spurious recursion effects, the
dynamics is evaluated up to a temporal cut-off of $\sim 2N/3$. We
checked that the precise cut-off is not relevant and that the plot
remains unchanged if $N$ is varied, provided $N\gg1$. All quantities are dimensionless.}
\label{quant}
\end{figure}

\begin{figure}[b]
\includegraphics[width=0.8\linewidth]{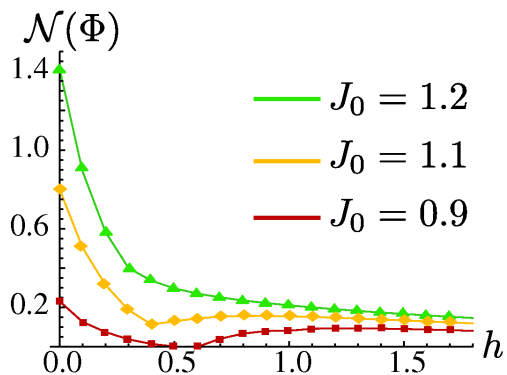}
\caption{(Color online) ${\cal N}(\Phi)$ versus $h/J$ for $h_0/J=0$
and three different values of $J_0/J$ for a chain of
$N=200$ spins. For $J_0/J{>}1$, the Markovianity point disappears. All quantities are dimensionless.
} \label{tutte}
\end{figure}

Other points in parameter space exist for which the model is
amenable to an exact analytic solution. However, in order to give
a complete overview of the behavior of ${\cal N}$, we first resort
to numerical techniques to solve our model. The results of such an
analysis are shown in Fig.~\ref{quant}, where the quantitative
degree of non-Markovianity is shown  as a function of $h/J$ (we set $h_0{=}0$) 
in the isotropic case and for equal intra-chain and
qubit-environment coupling strengths.

A highly non-trivial behavior followed by ${\cal N}(\Phi)$ is
revealed. The largest deviation from a Markovian dynamics is
achieved at $h/J=0$ while, for $0\leq h/J \leq 0.5$,
${\cal N}(\Phi)$ decreases monotonically to zero. By further increasing
$h$ we see that ${\cal N}(\Phi)$ achieves a very broad
maximum around the saturation point $h/J\simeq{1}$ of the
environmental chain. Finally, it goes to zero for $h/J\gg1$, as it
should be expected given that this situation corresponds to an
effective decoupling of the qubit from the
environment~\cite{ApollaroEtal10}. By generalizing our study to the case of $h_0/J\neq{0}$, we emphasize the presence of a Markovianity point
at $\delta{h}/J{=}1/2$. It turns out that this point separates two
regions in which the dynamics of the qubit (although being
non-Markovian in both cases) is completely different. For $\delta
h/J{\leq}1/2$, indeed,  the qubit tends toward a unique equilibrium
state at long times, irrespective of the initial condition; that
is, the trace distance goes to zero after some oscillations. For
larger detunings, on the other hand, the trace distance does not
decay to zero, implying that some information about the initial
state (and in particular, about the relative phase between its two
components) is trapped in the qubit.

The Markovianity point at $\delta h/J{=}{1}/{2}$ (with $J_0/J{=}1$) is
one of those points in parameter space for which the model can be
treated fully analytically~\cite{yueh04}. We have
$\Pi_0(t){=}{\cal J}_0(2t) \cos t{+}{\cal J}_1(2t) \sin t$ and
$\Delta_0(t){=}{\cal J}_1(2t) \cos t{-}{\cal J}_0(2t) \sin
t$~\cite{ApollaroEtal10}, so $f(t){=}{\cal J}_0^2(2t){+}{\cal
J}_1^2(2t)$ and $f'(t){=}{-}{2{\cal
J}_1^2(2t)}/{t}{<}0~\forall{t}$. As a consequence, the
corresponding non-Markovianity measure is always zero at this
point of the parameter space, which is a very interesting result
due to the relatively large $Q-\Gamma$ coupling strength. Such a
feature, in fact, would intuitively lead to exclude any
possibility of a forgetful dynamics undergone by the qubit. Yet,
this is not the case and a fully Markovian evolution is in order
under these working conditions. A deeper characterization of this
Markovian dynamical map is given in Sec.~\ref{caratterizzazione}.

As $J_0$ represents the energy scale of the qubit-environment
interaction, it is natural to expect that significant changes in
${\cal N}(\Phi)$ occur as this parameter varies. In particular, we find
that {\it i}) a Markovianity point with ${\cal N}(\Phi){=}0$ only
exists for $J_0/J{\le}1$ and {\it ii}) ${\cal N}(\Phi)$ tends progressively 
towards 
a monotonically decreasing
function of $\delta h$ if $J_0/J$ grows from $1$ to $\sqrt{2}$ [see Fig.~\ref{tutte}]. 
Strikingly, at $J_0/J{=}\sqrt{2}$ and $\delta h{=}0$,
the adopted measure of non-Markovianity diverges, as it can be
checked by using the analytic integrability of the qubit-chain
interaction at this point in the parameter space. Indeed, at
$J_0/J{=}\sqrt{2}$, we have $\Pi_0(t)={\cal J}_0(2t)$ and
$\sigma(t){=}-2\text{sgn}[{\cal J}_0(2t)]{\cal J}_1(2t)$.
Integrating over all the positive time intervals determined by means
of the usual chain rule we get ${\cal N}(\Phi){\rightarrow}\infty$,
thus witnessing a strong back-action of $\Gamma$ on the state of
the qubit. The divergence of ${\cal N}(\Phi)$ should not surprise
as it is common to other situations with spin-environments, such
as the so-called central-spin model where a single qubit is
simultaneously coupled to $N$ independent environmental spins via
Ising-like interactions (see Breuer {\it et al.} in
Ref.~\cite{breuer09}). We provide a physical
explanation for the enhanced non-Markovian nature of the qubit
dynamics simply by looking at the spectrum of the Hamiltonian
ruling the evolution of the qubit-chain system. For
$J_0/J{\ge}\sqrt{2}$, the spectrum of $\hat{\cal H}_\Gamma+\hat{\cal
H}_0$ exhibits a continuous spectrum (a {\it band} of extended
eigenstates) that is lower- and upper-bounded by two discrete
energy levels whose eigenstates are localized at the sites
occupied by the qubit and the first spin of $\Gamma$ \cite{nota}.
As a consequence, a certain amount of information remains trapped
into such a localized state, bouncing back and forth between the
qubit and the first spin and therefore mimicking a highly
non-Markovian dynamics characterized by strong back-action, so that ${\cal
N}(\Phi)$ diverges.

\begin{figure}[b]
\includegraphics[width=0.8\linewidth]{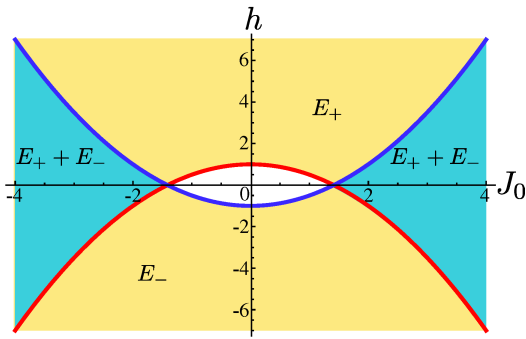}
\caption{(Color online) We call $E_+$ ($E_-$) a discrete level
lying above (below) the continuous energy band in the (single
particle) spectrum of the total Hamiltonian $\hat{\cal
H}_0+\hat{\cal H}_\Gamma$. The parabolae $h/J=\pm[1-(J_0/J)^2/2]$ divide
the $(h/J,J_0/J)$ parameter plane in three regions with one, two and
no localized state (light-colored, dark-colored and uncolored
region in the plane, respectively). All quantities are dimensionless.} \label{parabole}
\end{figure}

An analysis similar to the one performed just above 
allows us to obtain an intuition for the behavior of the measure of
non-Markovianity near this point. For $\delta h/J{>}{1}/{2}$ the spectrum of the
system shows one eigenenergy out of the band and its
corresponding eigenvector is localized around the site occupied by
$Q$. This can explain the information trapping that occurs for
$\delta h/J{>}1/2$. 
For definiteness, in what follows we report explicit results for the
case $h_0/J=0$. Therefore, from now on, we consider $\delta h/J \equiv
h/J$.

In order to determine the existence of a more general connection
between the emergence of localized eigenstates in the
system-environment spectrum and a point of zero-${\cal N}$ in the
qubit evolution, we analyze the $(h,J_0)$-plane to find out where
localized eigenstates appear and then evaluate the corresponding
degree of non-Markovianity. In doing this, we take advantage of
the fact that the Hamiltonian describing an environmental
$XX$-model  has the same single-particle energy spectrum as a
tight-binding model with an impurity. Following the approach given
in Ref.~\cite{pury91}, we deduce that, in the $(h/J,J_0/J)$ plane, the
parabolae $h/J=\pm[1-(J_0/J)^2/{2}]$ define regions with respectively
zero, one and two localized energy levels out of a
continuous-energy band (see Fig.~\ref{parabole}).

The central region with no localized energy state and the zones
with only one localized state (either a upper-lying or lower-lying
one with respect to the continuous band) correspond to ${\cal
N}(\Phi)\neq0$, although finite. The frontiers of such regions,
marked by the parabolae, give the limiting values of $(h/J,J_0/J)$ for
the appearance of the localized states and are such that ${\cal
N}(\Phi)=0$ only for $J_0/J\leq 1$, whereas for $1<J_0/J<\sqrt{2}$ the
measure of non-Markovianity is non-null and finite. Finally, the
regions with two localized states  and their frontier with the
no-localized state region have ${\cal
N}(\Phi){\rightarrow}\infty$, in line with the analytical results
discussed above.

In particular, we remark once again that the Markovianity points
(when they exist, i.e. for $J_0/J \leq 1$) stay on the parabolae;
that is, they are found to occur at the onset for the existence of
one discrete eigenstate outside the energy band. This kind of
state contains a spatially localized spin excitation with a
localization length that decreases with increasing $h$
\cite{nota}. Precisely at the border, the localization length
becomes as large as the length of the chain itself, so that all of
the environmental spins are involved in (or share excitation of)
the initial state. This can intuitively justify the fact that
${\cal N}$ is zero there. For $J_0/J >1$, on the other hand, the
first spin of the environment becomes more important for the
dynamics of the qubit (which is ``strongly coupled'' to it) and an
information exchange is always found to occur between them,
irrespectively of the existence of the discrete level. This
information re-flux becomes more and more pronounced with
increasing $J_0/J$ and decreasing $h/J$.

\begin{figure}[b]
\includegraphics[width=0.8\linewidth]{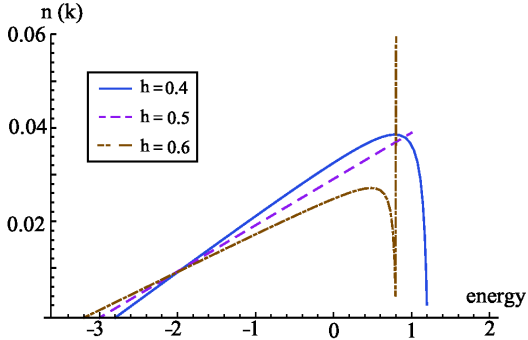}
\caption{(Color online) Average number of fermion excitations in
the initial state for a qubit-chain system with $N+1=50$, coupling
strength $J_0/J=1$ and three values of magnetic field around the
Markovianity point $(h/J=0.5, h_0/J=0)$. The initial state is always
taken to be the tensor product of an equatorial state of $Q$ and
the ground state of $\Gamma$.} \label{densita}
\end{figure}

To stress once more the close relationship between the
non-Markovianity measure and the properties of the overall
Hamiltonian $\hat{\cal{H}}_{\Gamma}+\hat{\cal{H}}_{0}$, we report
here the energy distribution of the excitations which are present
in the initial state of the system (given by the product of an
equatorial state for the qubit times the ground state of the
environment). These excitations are spin-less fermions of the
Jordan-Wigner type and the procedure to obtain them is the one
described, e.g., in Ref. \cite{apollaro08}. Fig. \ref{densita}
reports the average value of the excitation number in the initial
state vs single-particle-energy for various values of the magnetic
field. The plot shows what happens near a Markovianity point: the
energy distribution of initial-state excitations becomes flat
(i.e., structure-less) at the ${\cal N}=0$ point, while it shows a
maximum for lower values of $h$, corresponding to a finite value
of ${\cal N}$, and a spike for $h/J>0.5$, corresponding to the
discrete level giving rise to information trapping.

From this discussion we conclude that the measure of
non-Markovianity of the qubit dynamics is in fact a detector of
general  aspects of the full qubit+environment system and that
various features of ${\cal N}(\Phi)$ can be related to general
characteristics of the overall many-body problem described by the
full Hamiltonian model.

\section{Characterization of the point of zero-measure}
\label{caratterizzazione}

The occurrence of a null value of ${\cal N}(\Phi)$ at $h/J=1/2$ and
$J_0/J=1$ deserves a special attention. Naively, one could expect a
{\it special} behavior to occur at the chain saturation point
({\it i.e.} at $h/J=1$), where the intrinsic properties of the
environmental system are markedly different from the situation at
$h/J<1$. To the best of our knowledge, indeed, no significant
dynamical feature has been reported for the model under scrutiny
away from saturation.

\subsection{Characterization of the dynamical map: formal features and divisibility}

The aim of this Section is to characterize the dynamical map that
we obtained for the qubit under these conditions.
\begin{figure}[b]
{\bf (a)}\hskip4cm{\bf (b)}
\includegraphics[width=0.5\linewidth]{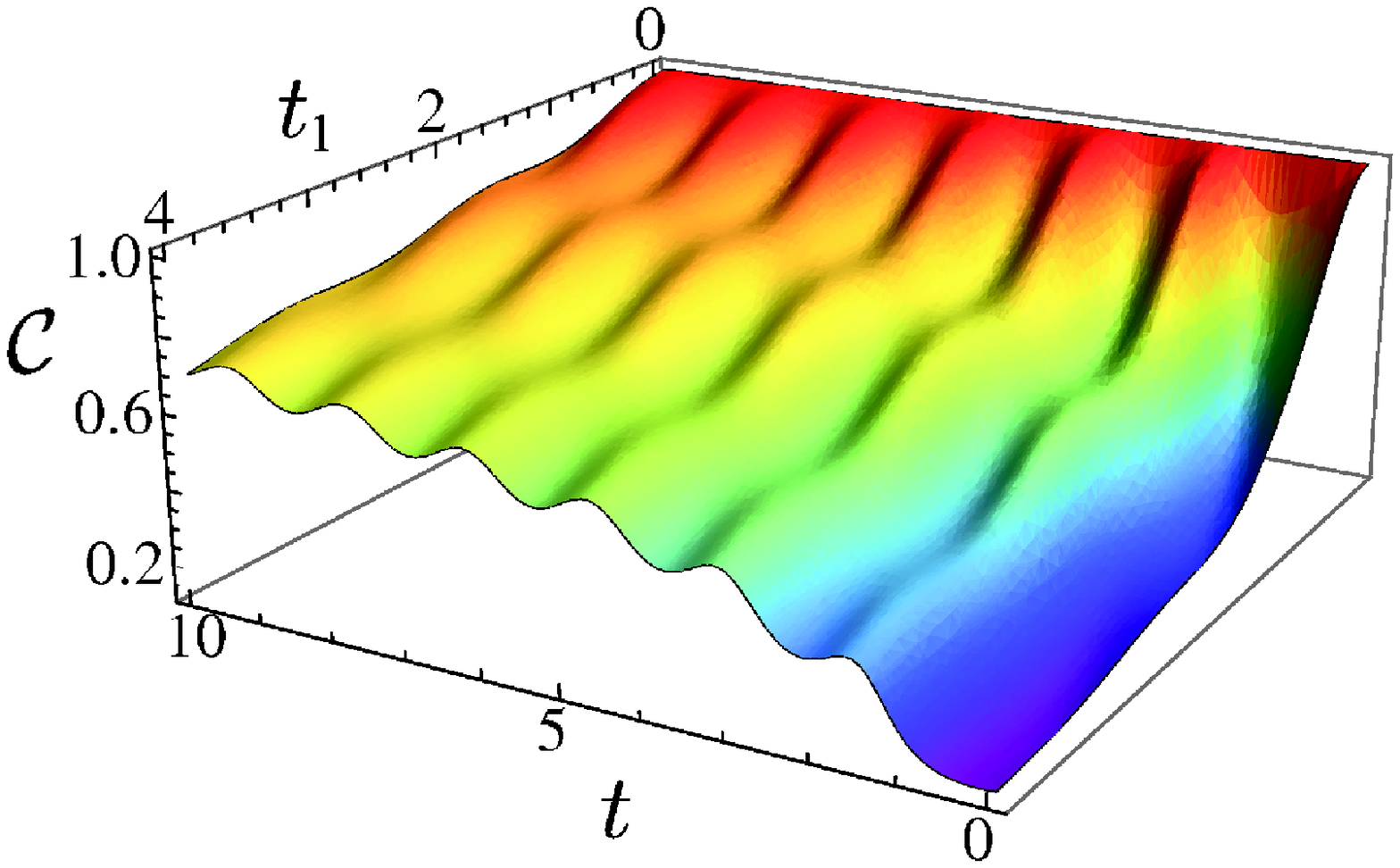}~~\includegraphics[width=0.45\linewidth]{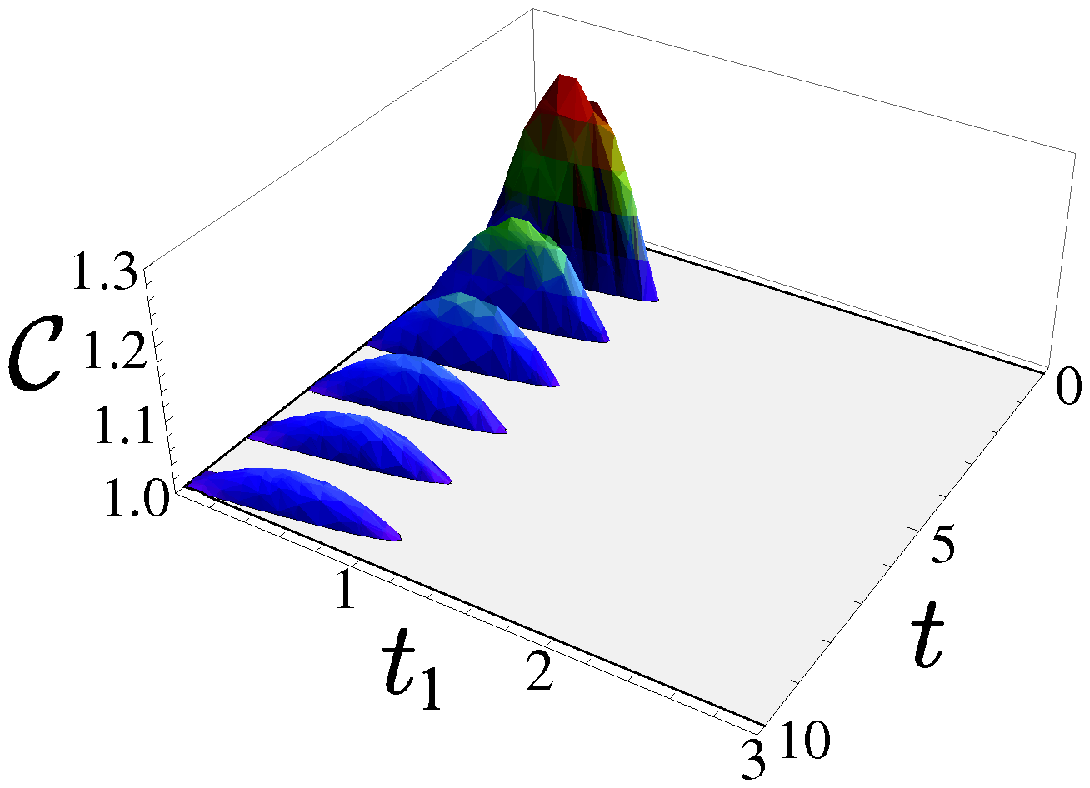}
\caption{(Color online) Divisibility condition ${\cal C}$ against
$t$ and $t_1$ for qubit $Q$ initially prepared in
$(\ket{0}_0+\ket{1}_0)/\sqrt 2$. Divisibility is guaranteed for
$0\le{\cal C}\le{1}$. In panel {\bf (a)} we have taken $J_0/J=1$
with $h_0/J=h/J=1/2$, while in {\bf (b)} it is $h/J=1.1$. All quantities are dimensionless.}
\label{divi}
\end{figure}
Thanks to the analytical solution for the case at hand provided in
Ref.~\cite{yueh04}, we can sum up the terms appearing in
Eq.~(\ref{dinamica}) and determine the complete density matrix of
$Q$. For $n=0,..,N$ we have
\begin{equation}
\Pi_{n}(t){=}\begin{cases}
(-1)^{\frac{n}{2}}[\sin(t){\cal J}_{n+1}(2t){+}\cos(t){\cal J}_{n}(2t)]&\text{$n$ even},\\
(-1)^{\frac{n+1}{2}}[\cos(t){\cal J}_{n+1}(2t){-}\sin(t){\cal J}_{n}(2t)]&\text{$n$ odd}
\end{cases}
\end{equation}
and
\begin{equation}
\Delta_{n}(t){=}\begin{cases}
(-1)^{\frac{n}{2}}[\cos(t){\cal J}_{n+1}(2t){-}\sin(t){\cal J}_{n}(2t)]&\!\text{$n$ even},\\
(-1)^{\frac{n+3}{2}}[\sin(t){\cal J}_{n+1}(2t){+}\cos(t){\cal J}_{n}(2t)]&\text{$n$ odd}.
\end{cases}
\end{equation}
The dynamical map transforms the
elements of the input density matrix $\rho_{ij}(0)~(i,j=0,1)$ as
\begin{equation}
\label{map2}
\begin{aligned}
\rho_{00}(t)&= f(t) \rho_{00}(0)+A_{00}^{11}(t),~~\rho_{01}(t)&=A_{01}^{01}(t)\rho_{01}(0),\\
\rho_{11}(t)&= A_{11}^{00}(t)+f(t) \rho_{11}(0),~~\rho_{10}(t)&=A_{10}^{10}(t)\rho_{10}(0),
\end{aligned}
\end{equation}
with $A_{00}^{11}(t){=}[1{-}f(t){+}g(t)]/{2}$,
$A_{11}^{00}(t){=}[1{-}f(t){-}g(t)]/{2}$ and
$A_{01}^{01}(t){=}\Pi_0(t){+}i \Delta_0(t){=}[A_{10}^{10}(t)]^*$,
and where $f(t){=}\Pi_0(t)^2{+}\Delta_0(t)^2$ has already been
defined in Sec.~\ref{modelmeasure}. We introduced here the
function $g(t)=\sum_{n=1}^{N}
[\Pi^2_n(t)+\Delta^2_n(t)]\mean{\hat\sigma^z_n}-\frac{1}{4}\sum_{\substack{n\neq{m}=1}}^N[\Pi_n(t)\Pi_m(t)+\Delta_n(t)\Delta_m(t)]g_{nm}$,
written in terms of the magnetization and the two-point
longitudinal correlation function
\begin{equation}
\begin{aligned}
\mean{\hat\sigma_n^z}&{=}1{-}\frac{2}{N+1}
\left(k_F-\frac{\cos[(k_F+1)\vartheta_n]\sin[k_F \vartheta_n]}{\sin\vartheta_n}\right),\\
g_{nm}&{\equiv}\mean{\hat P_{n}\hat P_m\hat\sigma_n^x\hat\sigma_m^x}=
\frac{\varphi_{n,k_F+1}\varphi_{m,k_F}-\varphi_{n,k_F}\varphi_{m,k_F+1}}
{2\left(\cos\vartheta_n-\cos\vartheta_m\right)},
\end{aligned}
\end{equation}
with $\varphi_{j,k}{=}\sqrt{{2}/{(N+1)}}\sin(j \vartheta_k)$,
${\vartheta_k={k\pi}/({N+1})}$, ${k=1,..,N}$ and $k_F$ being the
Fermi wave number, see Ref.~\cite{wonmin}.

The condition for divisibility stated in Ref.~\cite{breuer09}
implies the  existence of a completely positive dynamical map
$\Psi(t+t_1,t)$ such that, for two arbitrary instants of time $t$
and $t_1$, we have ${\Phi}(t+t_1,0)={\Psi}(t+t_1,t){\Phi}(t,0)$.
Here $\Phi(t,0)$ is the dynamical map in Eq.~(\ref{map2}). Any
dynamical map that is divisible according to the above definition
is Markovian. This implies that non-divisibility is a necessary
condition for memory-keeping effects in the evolution of a system.
A dynamical connection $\Psi(t+t_1,t)$ between the states
${\Phi}(t+t_1,0)\rho(0)$ and ${\Phi}(t,0)\rho(0)$ can be
straightforwardly found to be given by the map changing the
elements of the qubit state $\rho(t)$ at time $t$ into
\begin{equation}
\label{map3}
\begin{aligned}
\rho_{00}(t+t_1)&{=}\frac{f(t+t_1)}{f(t)}\rho_{00}(t){+}A_{00}^{11}(t+t_1){-}\frac{f(t+t_1)}{f(t)}A_{00}^{11}(t),\\
\rho_{11}(t+t_1)&{=}\frac{f(t+t_1)}{f(t)}\rho_{11}(t){+}A_{11}^{00}(t+t_1){-}\frac{f(t+t_1)}{f(t)}A_{11}^{00}(t),\\
\rho_{01}(t+t_1)&{=}\rho^*_{10}(t+t_1)=\frac{A_{01}^{01}(t+t_1)}{A_{01}^{01}(t)}\rho_{01}(t).
\end{aligned}
\end{equation}
Therefore, in order to ensure the divisibility of $\Phi(t+t_1,0)$, we
should  investigate the complete positivity of $\Psi(t+t_1,t)$. To this purpose we make use of the Choi-Jamiolkowski  isomorphism~\cite{CJ} and prove the complete positivity of $\Psi(t+t_1,t)$ 
by checking the non-negativity of $\left(\Psi(t+t_1,t)\otimes \mathbf{I}_{2}\right)\rho$, where 
$\rho$ is the density matrix of one of the Bell states~\cite{ruskai} and $\mathbf{I}_{2}$ is the identity map.
By choosing $\rho(t){=}\frac{1}{2}\left(\ket{00}+\ket{11}\right)\left(\bra{00}+\bra{11}\right)$,
the action of the map $\left(\Psi(t+t_1,t)\otimes \mathbf{I}_{2}\right)$ determines the following non-zero matrix elements (up to an irrelevant factor ${1}/{2}$) at time $t+t_1$
\begin{equation}
\begin{aligned}
&\rho_{00,00}(t+t_1,t){=}\frac{f(t+t_1)}{f(t)}{+}A_{00}^{11}(t+t_1){-}\frac{f(t+t_1)}{f(t)}A_{00}^{11}(t),\\
&\rho_{00,11}(t+t_1,t){=}A_{00}^{11}(t+t_1){-}\frac{f(t+t_1)}{f(t)}A_{00}^{11}(t),\\
&\rho_{11,00}(t+t_1,t){=}A_{11}^{00}(t+t_1){-}\frac{f(t+t_1)}{f(t)}A_{11}^{00}(t),\\
&\rho_{11,11}(t+t_1,t){=}\frac{f(t+t_1)}{f(t)}{+}A_{11}^{00}(t+t_1){-}\frac{f(t+t_1)}{f(t)}A_{11}^{00}(t),\\
&\rho_{01,01}(t+t_1,t){=}\rho^*_{01,01}(t+t_1,t)=\frac{A_{01}^{01}(t+t_1)}{A_{01}^{01}(t)}.\\
\end{aligned}
\end{equation}
Here, we have used $\rho_{i' j',i j}{=}\bra{i' i}\rho\ket{j' j}$ with the primed (unprimed) indeces corresponding to the evolving (non-evolving) qubit.  
The condition for positivity of the composite two-qubit map 
turns out to be equivalent to the positivity condition of the single-qubit one given in Eqs.~(\ref{map3}), which is in turn translated into the inequality
${0\le{\cal C}(t,t_1)\le{1}}$ with
\begin{equation}
{\cal C}(t,t_1)=4|\rho_{01}(t+t_1)|^2+[\rho_{00}(t+t_1)-\rho_{11}(t+t_1)]^2.
\end{equation}
In Fig.~\ref{divi} we show the typical behavior of ${\cal
C}(t,t_1)$ at the Markovianity point [panel {\bf (a)}] and away
from it [panel {\bf (b)}]. We have taken the qubit $Q$ as prepared
in $(\ket{0}_0+\ket{1}_0)/\sqrt 2$, which is a significant case as
equatorial states in the Bloch sphere are those optimizing the
calculation of ${\cal N}(\Phi)$. Although this is simply a
representative case, we have checked that for a uniform
distribution of random initial states of $Q$, no significant
quantitative deviations from the picture drawn here are observed.
Clearly, by moving away from $h/J{=}1/2$, temporal regions where
${\cal C}(t,t_1){>}1$ are achieved. This demonstrates, from a slightly different perspective, the
flexibility of the effective qubit evolution: a wide range of
dynamical situations is spanned, from fully forgetful to deeply
non-Markovian dynamics, strongly affected by the environmental
back-action. The kind of evolution of $Q$ can be determined by
tuning the parameters of the environment and its interaction with
it.

\subsection{Formal characterization of the channel through theoretical quantum process tomography}

We now turn our attention towards the formal characterization of
the channel achieved at the Markovianity point, so as to
qualitatively explain the reasons behind the nature of
the corresponding qubit evolution. We stress that this sort of
investigation is meaningful only at this specific point in
parameter space, where complete positivity is guaranteed. 
In principle, full information on the reduced dynamics of $Q$
could be gathered from the Kraus operators $\{\hat{K}_i\}$ such
that $\Phi(t,0)\rho{\equiv}\sum_i\hat{K}_i\varrho \hat{K}^\dag_i$
with $\rho$ the density matrix of the generic qubit and
$\sum_i\hat{K}_i^{\dag}\hat{K}_i{=}\hat\openone$. As their direct
computation is not possible due to the complications of the
$Q-\Gamma$ coupling, here we gain useful information on the
structure of the $\hat{K}_i$'s by means of the formal apparatus for
quantum process tomography~\cite{NC}, that we briefly remind here.

\begin{figure}[t]
\includegraphics[width=0.6\linewidth]{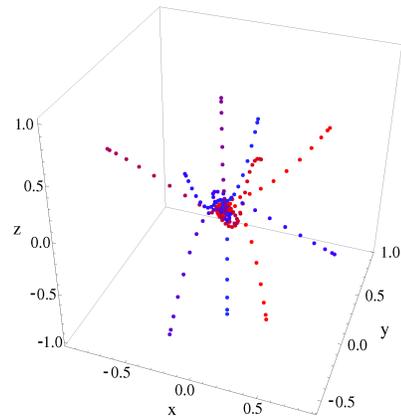}
\caption{(Color online) Dynamics of $10$ random pure initial
states of qubit $Q$ pictured in the Bloch sphere. Each dot shows
the qubit state  (evolved from the corresponding input state) at a
given instant of time. Different colors stand for different
initial states. Time increases as the dots move towards point
$(0,0,0)$, thus showing that $Q$ converges towards a state
close to a fully mixed one. } \label{spirale}
\end{figure}

The characterization of a dynamical map ${\Phi}$ reduces to the
determination of a complete set of orthogonal operators
$\{\hat{\cal K}_m\}$ over which one can perform the decomposition
$\hat{K}_i{=}\sum_m{e}_{im}\hat{\cal K}_m$ so as to get
\begin{equation}
{\Phi}(t,0)\varrho=\sum_{m,n}\chi_{mn}\hat{\cal K}_m\varrho\hat{\cal
K}^\dag_n ,
\end{equation}
where the {\it channel matrix} $\chi_{mn}{=}\sum_ie_{im}e^{*}_{in}$
has been introduced. This is a pragmatically very useful result as
it shows that it is sufficient to consider a fixed set of
operators, whose knowledge is enough to characterize a channel
through the matrix $\chi$. The action of $\Phi$ over a generic
element $\ket{n}\!\bra{m}$ of a basis in the space of the
$2{\times}{2}$ matrices (with $n,m{=}0,1$) can be determined from
a knowledge of the map ${\Phi}$ on the fixed set of states
$\ket{0},\ket{1},\ket{+}{=}(1/\sqrt 2)(\ket{0}{+}\ket{1})$ and
$\ket{+_y}{=}(1/\sqrt 2)(\ket{0}{+}i\ket{1})$ as follows
\begin{equation}
\begin{aligned}
{\cal E}(\ket{n}\bra{m})&={\cal E}(\ket{+}\bra{+})+i{\cal E}(\ket{+_y}\bra{+_y})\\
&-\frac{i+1}{2}[{\cal E}(\ket{n}\bra{n})+{\cal E}(\ket{m}\bra{m})].
\end{aligned}
\end{equation}
Therefore, each $\varrho_j=\ket{n}\bra{m}$ (with $j=1,..,4$) can
be found completely via state tomography of just four fixed
states. Clearly, ${\Phi}(\varrho_j)=\sum_k\lambda_{jk}\varrho_k$
as $\{\varrho_k\}$ form a basis. From the above
discussion we have
\begin{equation}
\label{determino}
\begin{aligned}
{\Phi}(t,0)\varrho_j{=}\sum_{m,n}\hat{\cal K}_m\varrho_j\hat{\cal K}^{\dag}_n
\chi_{mn}{=}\sum_{m,n,k}\beta^{mn}_{jk}\varrho_k\chi_{mn}{=}\sum_k\lambda_{jk}\varrho_k,
\end{aligned}
\end{equation}
where we have defined $\hat{\cal K}_m\varrho_j\hat{\cal K}^\dag_n=\sum_k\beta^{mn}_{jk}\varrho_k$ so that we can write
\begin{equation}
\label{beta}
{\lambda_{jk}=\sum_{m,n}\beta^{mn}_{jk}\chi_{mn}}.
\end{equation}
The complex tensor $\beta^{mn}_{jk}$ is set once we make a choice
for $\{\hat{{\cal K}}_i\}$ and the $\lambda_{jk}$'s are determined
from a knowledge of ${\Phi}(\varrho_j)$. By inverting
Eq.~(\ref{beta}), we can determine the channel matrix $\chi$ and
characterize the map. Let $\hat{\cal U}^\dag$ be the operator diagonalizing the channel
matrix. It is straightforward to prove that, if $D_i$ are the
elements of the diagonal matrix $\hat{\cal U}^\dag\chi{\hat{\cal
U}}$, then $e_{im}=\sqrt{D_i}\hat{U}_{mi}$ so that
\begin{equation}
\hat{K}_{i}=\sqrt{D_i}\sum_j\hat{\cal U}_{ji}\hat{\cal K}_j.
\end{equation}
This apparatus can be applied to the problem at hand. To this end,
we numerically determine the evolution of the four qubit states
$\ket{0}$, $\ket{1}$, $\ket{+}$ and $\ket{+_y}$ at $h/J{=}1/2$,
$h_0{=}0$ and in the uniform-coupling case for $N{+}1{=}150$. As
the time-scale within which an excitation travels back and forth
to $Q$ scales as $N$~\cite{ApollaroEtal10}, in order to avoid
recurrence effects we limit the time-window of our analysis to
$[0,2N/3]{=}[0,100]$, which we chop into small intervals of
amplitude $0.2$. At each instant of time in such a temporal
partition, we evaluate the evolution of the four probing states
given above. This is the basis for the reconstruction of the Kraus
operators, which is performed as described above. The results of
our numerical calculations are four Kraus operators, whose form
depends on the instant of time at which the dynamics is evaluated.
In principle, an analytic form of such operators can be given.
Upon inspection, one can see that the action of the channel
embodied by $\Gamma$ over the four probing qubit states can be
summarized as follows
\begin{equation}
\begin{aligned}
&\ket{0}{\rightarrow}\varrho_0{=}
\begin{pmatrix}
\alpha&0\\
0&1-\alpha
\end{pmatrix},
\ket{+}{\rightarrow}\varrho_+{=}
\begin{pmatrix}
a&b+i c\\
b-i c&1-a
\end{pmatrix},\\
&\ket{1}{\rightarrow}\varrho_1{=}
\begin{pmatrix}
\beta&0\\
0&1-\beta
\end{pmatrix}\!,
\ket{+_y}{\rightarrow}\varrho_{+_y}{=}
\begin{pmatrix}
a&-c-ib\\
-c+ib&1-a
\end{pmatrix}
\end{aligned}
\end{equation}
with $a,\alpha,\beta\in[0,1]$ and $b,c\in\mathbb{C}$. The form of
the corresponding  analytical Kraus operator can then be found,
although their cumbersome nature makes them unsuitable to be
reported here. Despite the lack of analytic formulae, useful information can be
gathered from this study. Fig.~\ref{spirale} shows the {\it
dynamics} of 10 random pure input states of the qubit. Each dot
represents the state of the qubit at time $t$ in the space of the
density matrices. Colors identify the evolved states of a given
input one. As time increases, {\it i.e.} as we follow  the points
of the same color from the outer to the inner region of the plot,
the Bloch vector of each initial pure state {\it spirals}
converging towards the center. That is, the state becomes close to
a maximally mixed state.

We should mention that, for any fixed $N$, this does not exactly
occur as the final state of $Q$ lies slightly off-set along the
$z$-axis and does not describe a perfect statistical mixture. We
ascribe such an off-set to the finite length of the chain being
studied and the existence of a sort of effective magnetic field
(scaling as $1/N$) acting on $Q$ and induced by its coupling with
the environmental chain, which determines the residual
polarization of the qubit. Numerical evidences suggest that, in the
thermodynamical limit, such a residual polarization effectively
disappears and the final state of the evolution is $\openone/2$.

\begin{figure}[b]
{\bf (a)}\hskip4cm{\bf (b)}\\
\includegraphics[width=0.51\linewidth]{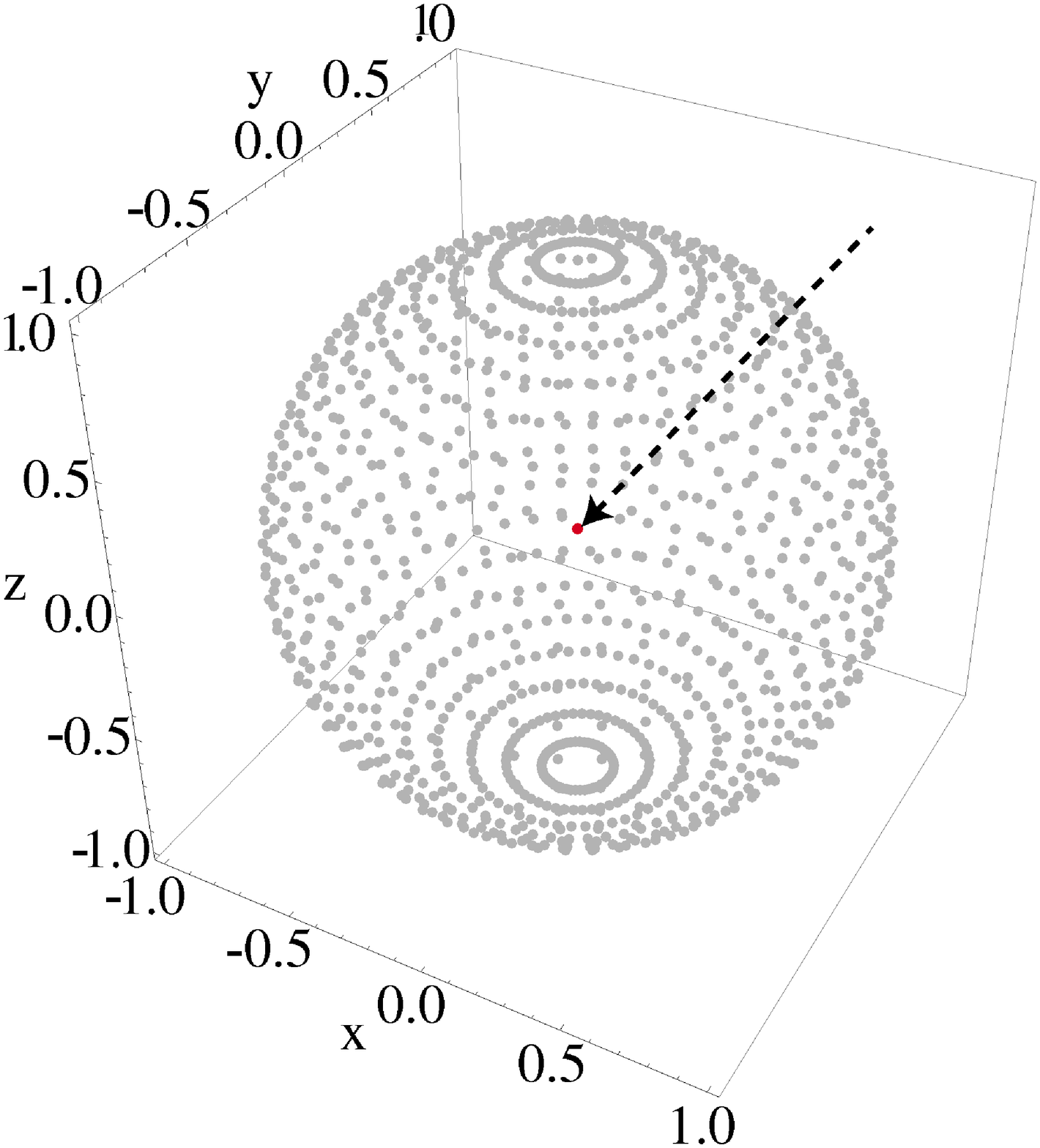}\includegraphics[width=0.5\linewidth]{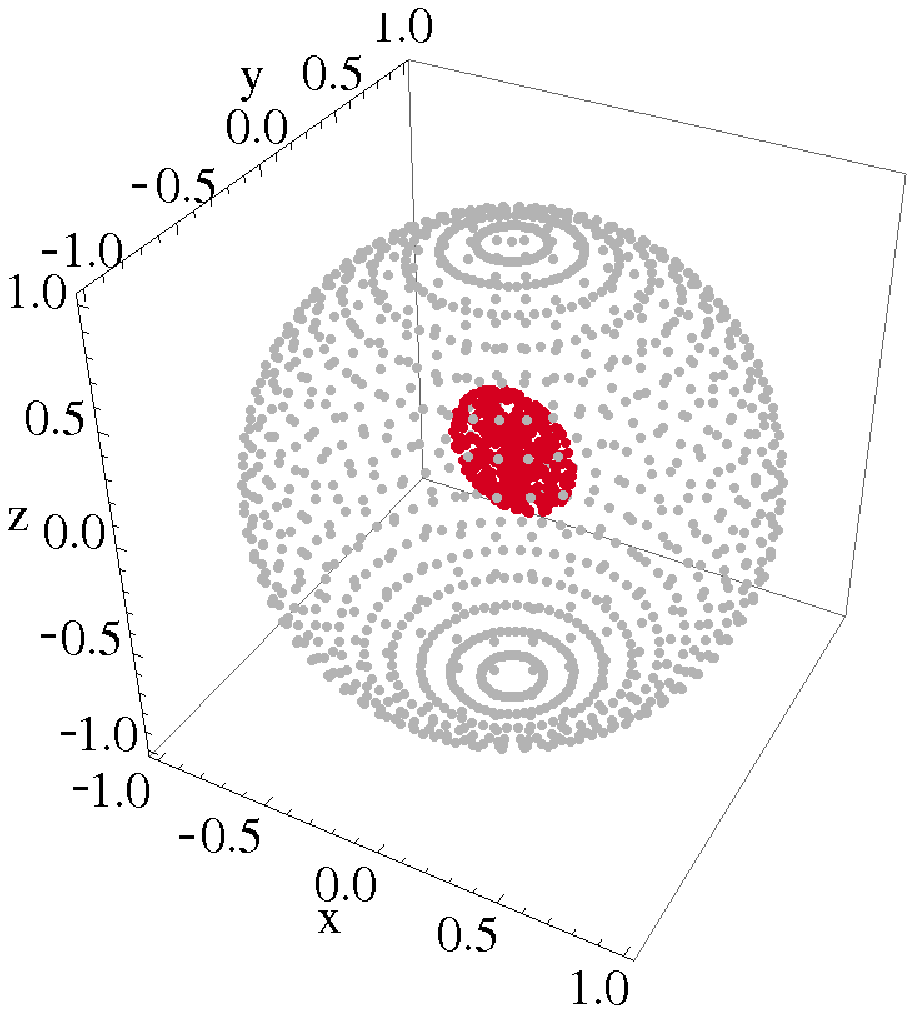}
\caption{(Color online) {\bf (a)} Distribution of density matrices
evolved  according to the dynamics induced by a quantum channel
corresponding to $h/J=1/2$. We consider a sample of $500$ random
initial states. Regardless of the initial density matrix, the
final state is the same, thus resulting in a zero-volume state
distribution. This is no longer the case for $h/J=0.6$ [panel {\bf
(b)}], where a finite-volume distribution of final states is
found. } \label{bloch}
\end{figure}

As discussed in the previous Section, from the study of the trace
distance ${\cal D}[\rho^{(1)}(t),\rho^{(2)}(t)]$ at the basis of
our chosen measure of  non-Markovianity, one can infer an
interesting general trend: for values of $h/J \leq 1/2$, ${\cal D}$
generally tends to zero at long evolution times. 
In order to put such an effect in context, here we study the
long-time distribution  of density matrices resulting from the
evolution of a large sample of random input states according to
the channel identified through quantum process tomography.
Fig.~\ref{bloch} {\bf (a)} shows the agreement between the
effective description provided here and the true physical effect
induced on the state of $Q$ by $\Gamma$.
The red dot close to the center of the sphere shows the ensemble
of $500$ random input states evolved until a final time equal to
$100$. Clearly, there is a single fixed point of the map towards
which any input state converges after a sufficient amount of time.

As anticipated, by repeating the above analysis for $h/J{>}1/2$, we
get a distribution of long-time evolved output states which is
much more widespread, even at very small deviations from the
Markovian case. An example is given in Fig.~\ref{bloch} {\bf (b)}
for $h/J{=}0.6$. At long interaction times, the channel does not bring
its input states to the same unique state, but a
finite-dimensional subspace of asymptotic states exists.

\begin{figure*}[t]
{\bf (a)}\hskip6cm{\bf (b)}\hskip6cm{\bf (c)}
\includegraphics[width=0.9\linewidth]{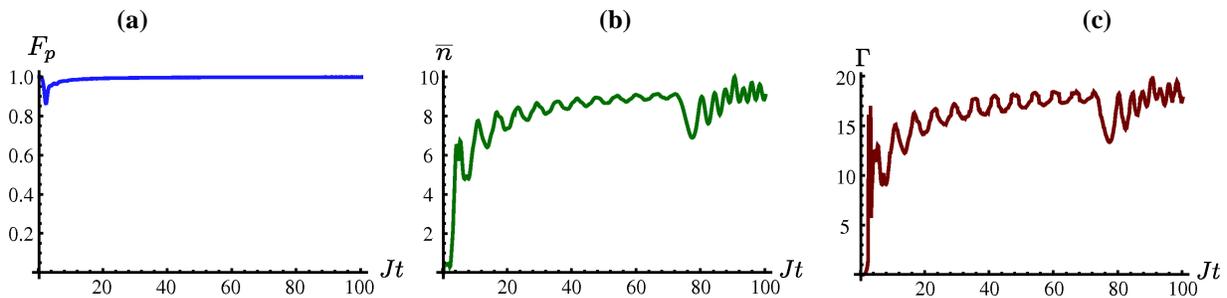}
\caption{(Color online) {\bf (a)}: Optimized quantum process
fidelity between a generalized amplitude damping channel and the
dynamical map here under scrutiny. The process fidelity has been
maximized with respect to the parameters of the generalized
amplitude damping channel at each instant of time. In panel {\bf
(b)} and {\bf (c)} we show the time-behavior of $\mu$ and
$\Gamma$. An analogous behavior is found for $\gamma$. The
time-dependence of the last two parameters, and in particular
their positivity, are fully in line with the Markovian nature of
the qubit effective dynamics.} \label{parametri}
\end{figure*}

Going back to our  attempt of describing the dynamics as an
effective channel, we have considered the case of a qubit evolving
under a Markovian channel resulting from the action of a
generalized amplitude damping channel and an independent dephasing
mechanism. The output state for such a process can be written as
\begin{widetext}
\begin{equation}
\label{fic}
\rho(t)=
\begin{pmatrix}
\rho_{00}(0)e^{-2\tilde\gamma(2\tilde\mu+1)t}+\frac{\tilde\mu+1}{2\tilde\mu+1}
[1-e^{-2\tilde\gamma(2\tilde\mu+1)t}]&\rho_{01}(0)e^{-[2\tilde\Gamma+
\tilde\gamma(2\tilde\mu+1)]t}\\
\rho_{10}(0)e^{-[2\tilde\Gamma+\tilde\gamma(2\tilde\mu+1)]t}&\frac{\tilde\mu}{2\tilde\mu+1}
+[\frac{\tilde\mu+1}{2\tilde\mu+1}-\rho_{00}(0)]e^{-2\tilde\gamma(2\tilde\mu+1)t}
\end{pmatrix}
\end{equation}
\end{widetext}
with $\tilde\mu$ being the mean occupation number of the thermal
environment responsible for the generalized amplitude damping,
$\tilde\gamma$ the rate of amplitude damping and $\tilde\Gamma$
the rate of dephasing. From this expression, one can determine the
matrix of the corresponding process, to be compared to the matrix
of the process associated with the qubit evolution induced by its
coupling to the chain determined by the quantum process tomography
machinery. Calling $\chi(t)$ such matrix evaluated at time $t$ and
$\chi_{c}$ the matrix of the process embodied by Eq.~(\ref{fic}),
the similarity between such two processes, as time passes, is
determined by the process fidelity
\begin{equation}
F_{p}=\left[\text{Tr}(\sqrt{\sqrt{\chi(t)}\chi_c\sqrt{\chi(t)}})\right]^2.
\end{equation}
In calculating this, we have considered scaled
parameters $\mu=\tilde\mu{t}$, $\gamma=\tilde\gamma t$ and
$\Gamma=\tilde\Gamma t$ and optimized $F_p$ against them.
Therefore, time-dependent values of such quantities are retained
at each instance of $Jt$ for the maximization of the process
fidelity, which is always at least $90\%$, as shown in
Fig.~\ref{parametri}.
This gives strong numerical indications that our process is an
homogenizing, time-dependent Markovian channel.

\section{Conclusions}
\label{conc}

We have studied the dynamics of a qubit interacting via
energy-exchange mechanisms with a spin chain under the viewpoint of a
recently proposed measure for non-Markovianity~\cite{breuer09}. We
have provided an extensive characterization of the effective
open-system evolution experienced by the qubit, often giving fully
analytical expressions for the measure of non-Markovianity and an
operative recipe for its experimental determination under certain
conditions. Our investigation has allowed us  to reveal various
features of the non-Markovianity measure and to relate them to
general properties of the overall $Q-\Gamma$ system. In
particular, we have focused on the existence of unexpected points in
the parameter space of the model, where the chosen measure of
non-Markovianity is exactly null, so that the qubit evolution
corresponds to a Markovian dynamical map. By using techniques
typical of quantum process tomography, we have been able to shed
light onto the reasons behind the forgetful nature of such {\it
peculiar} working point, showing, in particular, that the
zero-measure dynamical map has a single fixed point corresponding
to an almost completely mixed qubit state. We have found an excellent
agreement between the effective Markovian map and a generalized
amplitude damping channel with an additional (independent)
dephasing mechanism. The comparison between the two descriptions
allows us to determine the behavior of the effective dephasing and
damping rates, which are time-dependent yet positive, in agreement
with the Markovian nature of the dynamics at hand~\cite{breuer09}.

The richness of the qubit open-system dynamics is very
interesting.  We believe that a more extensive exploration of the
possibilities offered by the tunability of the degree of
Markovianity in such system should be performed so as to
understand if a proper and arbitrary ``guidance" of the qubit
state via the sole manipulation of the properties of the
environmental chain is in order. This topic is the focus of our
currently ongoing work which will be reported
elsewhere~\cite{futuro}.

\section{Acknowledgements}
FP acknowledges useful discussions with J. Piilo. TJGA thanks Ruggero Vaia for useful discussions and the School of Mathematics and Physics at Queen's University Belfast for hospitality. CDF is supported
by the Irish Research Council for Science, Engineering and Technology. TJGA acknowledges support of the Italian Ministry of Education, University, and Research in the framework of the 2008 PRIN program (Contract No. 2008PARRTS003). MP is supported by EPSRC (EP/G004579/1). MP and FP
acknowledge support by the British Council/MIUR British-Italian Partnership Programme 2009-2010.


\begin{thebibliography}{99}

\bibitem{wolf08}
M. M. Wolf, J. Eisert, T. S. Cubitt, and J. I. Cirac, Phys. Rev. Lett. {\bf 101}, 150402 (2008).

\bibitem{rivas09} A. Rivas, S. F. Huelga, and M. B. Plenio, Phys. Rev.
Lett. {\bf 105}, 050403 (2010).

\bibitem{breuer09}
H.-P. Breuer, E. M. Laine, and J. Piilo, Phys. Rev. Lett. {\bf
103}, 210401 (2009); E.-M. Laine, J. Piilo, and H.-P. Breuer,
Phys. Rev. A {\bf 81}, 062115 (2010).

\bibitem{martinis}
R.W. Simmonds {\it et al}., Phys. Rev. Lett. {\bf 93}, 077003
(2004); A. Lupascu {\it et al}.,  Phys. Rev. B {\bf 80}, 172506
(2009); Y. Shalibo {\it et al}., Phys. Rev. Lett. {\bf 105},
177001 (2010); E. Paladino, L. Faoro, G. Falci, and R. Fazio,
{\it ibid.} {\bf 88}, 228304 (2002); A. Shnirman, G. Sch\"on,
I. Martin, and Y. Makhlin, {\it ibid.} {\bf 94}, 127002
(2005);  G. Falci, A. D'Arrigo, A. Mastellone, and E. Paladino,
{\it ibid.} {\bf 94}, 167002 (2005);  Y. M. Galperin, B. L.
Altshuler, J. Bergli, and D. V. Shantsev, {\it ibid.} {\bf
96}, 097009 (2006); L. Faoro and L. B. Ioffe, {\it ibid.}
{\bf 96}, 047001 (2006); E. Paladino, M. Sassetti, G. Falci, and
U. Weiss, Phys. Rev. B {\bf 77}, 041303 (2008); J. Bergli, Y. M.
Galperin, B. L. Altshuler, New J. Phys. {\bf 11}, 025002
(2009); M. Constantin, C. C. Yu, and J. M. Martinis, Phys. Rev. B
{\bf 79}, 094520 (2009).

\bibitem{sun09} X.-M. Lu, X. Wang, and C. P. Sun, Phys. Rev. A {\bf 82}, 042103 (2010) .

\bibitem{boseetal}
S. Bose, Contemp. Phys. 48, 13 (2007); S. Bose, Phys. Rev. Lett.
{\bf 91}, 207901 (2003); M. Christandl {\it et al.}, Phys. Rev. A
71, 032312 (2005); L. Campos Venuti, C. Degli Esposti Boschi, and
M. Roncaglia, Phys. Rev. Lett. {\bf 99}, 060401 (2007); C. Di
Franco, M. Paternostro, M. S. Kim, Phys. Rev. Lett. {\bf 101},
230502 (2008); F. Plastina and T. J. G. Apollaro, {\it ibid.}
{\bf 99}, 177210 (2007); G. Gualdi, V. Kostak, I. Marzoli, and P.
Tombesi, Phys. Rev. A {\bf 78}, 022325 (2008).

\bibitem{rmp}
L. Amico, R. Fazio, A. Osterloh, and V. Vedral, Rev. Mod. Phys.
{\bf 80}, 517 (2008).

\bibitem{cinesi} Z. Y. Xu, W. L. Yang, and M. Feng, Phys. Rev. A {\bf 81}, 044105 (2010).

\bibitem{ApollaroEtal10}
T. J. G. Apollaro, A. Cuccoli, C. Di Franco, M. Paternostro, F. Plastina, and P. Verrucchi, New J. Phys. {\bf 12}, 083046 (2010).

\bibitem{DiFrancoEtal07} C. Di Franco, M. Paternostro, G. M. Palma, and M. S. Kim,
Phys. Rev. A {\bf 76}, 042316  (2007); C. Di Franco, M. Paternostro, and G. M. Palma,
Int. J. Quant. Inf. {\bf 6}, Supp. 1, 659 (2008).

\bibitem{statetomo} A. G. White {\it et al.}, Phys. Rev. Lett. {\bf 83}, 3103 (1999); D.
James {\it et al.}, Phys. Rev. A {\bf 64}, 052312 (2001).

\bibitem{DiFrancoEtal08} C. Di Franco, M. Paternostro, and M. S. Kim, Phys. Rev. Lett. {\bf 102}, 187203 (2009).


\bibitem{NC} M. A. Nielsen and I. L. Chuang, Quantum Computation and Quantum Information (Cambridge University Press, 2000);
I. L. Chuang and M. A. Nielsen, J. Mod. Opt. {\bf 44}, 2455 (1997).

\bibitem{yueh04}
W.-C. Yueh, Applied Mathematics E-Notes {\bf 5}, 66-74 (2005).

\bibitem{nota}
The form of such localized levels has been described by, e.g., L.
F. Santos, G. Rigolin, and C. O. Escobar, Phys. Rev. A {\bf 69},
042304 (2004); L. F. Santos and G. Rigolin, Phys. Rev. A {\bf 71},
032321 (2005); T. J. G. Apollaro and F. Plastina, Phys. Rev. A
{\bf 74}, 062316 (2006).

\bibitem{pury91}
P. A. Pury and S. A. Cannas, J. Phys. A {\bf 24}, L1405-L1414 (1991).

\bibitem{apollaro08}
T. J. Apollaro, A. Cuccoli, A. Fubini, F. Plastina, and P. Verrucchi, Phys. Rev. A {\bf 77}, 062314
(2008); L. Banchi, T. J. G. Apollaro, A. Cuccoli, R. Vaia, and P. Verrucchi, {\it ibid.} {\bf 82}, 052321 (2010).

\bibitem{CJ} A. Jamiolkowski, Rep. Math. Phys. {\bf 3}, 275 (1972); M.-D. 
Choi, Lin. Alg. and Appl. {\bf 10}, 285 (1975).

\bibitem{ruskai}
M. B. Ruskai, S. Szarek, and E. Werner, Lin. Alg. Appl. {\bf 347}, 159-187 (2002).

\bibitem{wonmin}
W. Son, L. Amico, F. Plastina, and V. Vedral, Phys. Rev. A {\bf 79}, 022302 (2009).

\bibitem{futuro} T. J. G. Apollaro {\it et al.} (work in progress).

\end{thebibliography}
\end{document}